\documentstyle[twocolumn,epsfig,seceq]{jpsj}
\title{Ground State Phase Diagram of 2D Electrons in High Magnetic Field}
\author{Naokazu {\sc Shibata} and Daijiro {\sc Yoshioka}}
\inst{
Department of Basic Science, University of Tokyo,
Komaba 3-8-1 Meguro-ku, Tokyo 153-8902, Japan
}
\recdate{October 23, 2002}

\abst{
The ground state of 2D electrons in high magnetic field
is studied by the density matrix renormalization group method.
The ground state energy, excitation gap, and 
pair correlation functions are systematically calculated 
at various fillings in the lowest and the second lowest Landau levels.
The ground state phase diagram, which consists of
incompressible liquid state, compressible liquid state, stripe state, 
pairing state, and Wigner crystal is determined.
}

\kword{fractional quantum Hall effect, stripe, pairing, Wigner crystal,  
ground state, phase diagram, two dimension, density matrix,
renormalization group}

\def\runtitle{Ground State Phase Diagram of 2D Electrons 
in High Magnetic Field}
\def\runauthor{Naokazu {\sc Shibata} and Daijiro {\sc Yoshioka}}

\begin{document}
\sloppy
\maketitle

\section{Introduction}
Electrons in high magnetic field in two dimensional systems  
exhibit various interesting phenomena as a typical 
many body interacting system.  
In two dimensional systems, kinetic energy of electrons is completely
quenched by perpendicular magnetic field and remaining macroscopic
degeneracy is lifted by Coulomb interaction.
This is the origin of many exotic behaviors of quantum Hall systems
and the reason why much effort has been needed to understand 
these systems.

As typical ground states in strong magnetic field,
Laughlin state and Wigner crystal have attracted much attention.
Laughlin state is a quantum liquid with an excitation gap,
which is an exact ground state for strong short range repulsive 
interaction at $\nu=1/(2m+1)$\cite{Lagh}.
On the other hand, Wigner crystal is the ground state of classical 
point particles realized in the limit of strong magnetic field. 
Their relative stability depends 
on the ratio between the magnetic length $\ell=(\hbar c/eB)^{1/2}$,
which is the length scale of one particle wave function,
and the mean distance between the electrons $\sim n^{-1/2}$ 
given by the density $n$ of two-dimensional electrons.
Thus the determination of the ground state phase diagram with respect 
to the filling factor $\nu = 2\pi \ell^2 n$ is an 
interesting issue for understanding the competition between 
completely different electronic states.

Experimentally, clear fractional quantum Hall effect has been observed 
at $\nu=1/3$\cite{exp3}, which is naturally explained by Laughlin 
state\cite{Lagh}, and exact numerical diagonalizations  
confirmed its existence in the lowest Landau level\cite{yoshi}.
The formation of Wigner crystal 
has been supported by experiments on transport properties
at low fillings $\nu  \stackrel{<}{_\sim} 1/5$, where 
observed insulating behavior is explained by the pinning
of Wigner crystal by residual potential fluctuations\cite{exp5,exp5n}.

With decreasing magnetic field,  
part of electrons occupy higher Landau levels.
In higher Landau levels, one particle wave function extends 
over space with oscillations, 
and this change modifies effective interaction 
between the electrons in each Landau level. 
Since extended one particle wave function generates long range
exchange interaction, Laughlin state stabilized by strong short range 
interaction will be unstable. 
Instead, long range interaction favors charge density waves (CDW's).
Within a Hartree-Fock approximation, Koulakov {\it et al}. showed 
that CDW's called 'stripe' and 'bubble' are stable against Laughlin 
state in higher Landau levels with index $N\ge 2$ \cite{Kou1,Kou2}. 
After their calculations, anisotropic resistivity 
consistent with uniaxial electronic state of stripe order has been
experimentally observed\cite{Lill,Du,Coop}.

In the second lowest Landau level, more interesting ground states
are expected. At $\nu_N=1/2$, where $\nu_N$ is the
filling factor of partially filled Landau level, BCS type pairing state has 
been proposed\cite{RezHal,grei,Moor} to explain even denominator 
fractional quantum 
Hall effect\cite{exp52}. 
Around $\nu_N=0.4$ and $\nu_N=0.24$, reentrant integer quantum Hall effects 
have been observed\cite{exp1}. At $\nu_N=1/3$ and $1/5$, 
fractional quantum Hall effects have been observed\cite{exp52,exp1}. 
The origin of these phenomena is still unclear and systematic study is needed.

In order to identify diverse ground states and determine the ground 
state phase diagram,
we have to solve quantum many body problem of Coulomb interaction.
In the present study we use the density matrix renormalization group 
(DMRG) algorithm\cite{DMRG,Shib} to deal with large size 
systems extending limitation of exact diagonalizations,
and systematically calculate the ground state wave function
and excitation energies for the lowest and the second lowest Landau levels. 
Based on the pair correlation functions and the excitation energies,
we determine the ground state phase diagram,
which consists of incompressible liquid state, compressible liquid
state, stripe state, paring state, and Wigner crystal.

\section{Model and Method}
The Hamiltonian for electrons in Landau levels
contains only Coulomb interaction. After the projection 
onto the Landau level with index $N$, the Coulomb interaction is written as
\begin{equation}
H=\sum_{i<j} \sum_{\bf q} e^{-q^2/2} \left[ L_{N}(q^2/2) \right] ^2 V(q) 
e^{i{\bf q} \cdot ({\bf R}_i-{\bf R}_j)} ,
\label{Coulomb}
\end{equation}
where ${\bf R}_i$ is the guiding center coordinate of the
$i$th electron, which satisfies the commutation relation,
$[{R}_{j}^x,{R}_{k}^y]=i\ell^2\delta_{jk}$,
$L_{N}(x)$ are the Laguerre polynomials,
and $V(q) =2\pi e^2/\varepsilon q$ is the Fourier transform of the
Coulomb interaction. 
The magnetic length $\ell$ is set to be 1,
and we take $e^2/\varepsilon \ell$ as units of energy scale.
We omit the component at $q=0$, which is canceled by 
uniform positive background charge.
We assume completely spin polarized ground state
and neglect the electrons in fully occupied Landau levels.
We also assume that the width of the wave function perpendicular to the two 
dimensional plane is sufficiently small compared with $\ell$.

In order to obtain the ground state of the Hamiltonian, we 
employ the DMRG method\cite{DMRG,Shib}.
This method enables us to calculate ground state wave function 
of large systems with controlled accuracy. 
The algorithm of this method is summarized as follows:
We start from a small system, i.e. a system consisting of
only four local orbitals. We divide the system into two blocks, 
and add new orbitals at the end of two blocks to expand the blocks.
We then calculate the ground state wave function of the total system
and obtain the density matrix of each expanded blocks.
We then restrict the basis states in the expanded blocks
by keeping only eigenstates of large eigenvalues of the
density matrix. We add new orbitals again and 
repeat the above procedure until we get desired size of system.
We then use the finite system 
algorithm of the DMRG to improve the ground state wave function. 
After several sweeps, we obtain the most optimal ground state within the 
restricted number of basis states.
The error due to the truncation of basis states is 
estimated from the eigenvalues of the density matrix,
and the accuracy of the wave function is systematically improved
by increasing the number of basis states kept in the blocks.

\section{Results}
In the following, we calculate ground state wave function of the 
Hamiltonian for Landau levels of $N=0$ and $1$.
We study various size of systems with up to 25 electrons
in the unit cell of $L_x\times L_y$ with periodic boundary 
conditions in both $x$ and $y$ directions. 
We choose the aspect ratio $L_x/L_y$ by the position of
the energy minimum with respect to $L_x/L_y$ 
in order to avoid artificial determination of the periodicity
of CDW's.
The truncation error in the DMRG calculation is
typically $10^{-4}$ for 25 electrons with 180 states in each blocks.
The existing results of exact diagonalizations are 
completely reproduced within the truncation error.
Since the present Hamiltonian has the particle-hole symmetry, 
we only consider the case of $\nu_N \le 1/2$.

\subsection{Compressible liquid at $\nu=1/2$}
We first investigate the ground state at $\nu=1/2$ in the lowest 
Landau level. 
Since the denominator of $\nu$ is even and the filling factor is 
large, the ground state is neither Laughlin state nor Wigner crystal.
Based on the composite fermion theory, the flux attached to the electrons
completely cancels external magnetic field within the mean field 
level and Fermi liquid ground state is expected\cite{jain,Halp}.
Indeed, experiments show saturation in
longitudinal resistivity at low temperatures\cite{exp2}, 
which is consistent with Fermi liquid like compressible ground state. 

In order to confirm the compressible ground state, we first analyze size 
dependence of the excitation gap. The obtained results for systems  
with up to 25 electrons are shown in Fig.~1. 
Although the excitation gap does not follow single scaling 
function, all the results are smaller than the upper bound 
shown by the dashed line, which is proportional to $N_s^{-1/2}$,
where $N_s$ is the number of one-particle states in the unit cell.
We thus expect vanishing of the excitation gap in the thermodynamic limit. 
We find small excitation gap at small number of electrons.
Such vanishing excitation gap appears for free electrons in 
the square unit cell at $N_e=2$,3,4,6,7,8,10,11,12,14,15,$...$,
with periodic boundary conditions.
This means large excitation gap appears only at $N_e=1$,5,9,13,21,25,$...$,
where electrons occupy all the degenerate states forming
closed shell. The present result of interacting system also shows
a shell structure but it is not the same to that of free 
electrons in contrast to the results in the spherical geometry,
where the same shell structure is obtained\cite{Onod}. Nevertheless, 
the estimated effective mass $1/m^*$ of the composite fermion determined
from the value of the gap at $N_e=9$ and $25$ 
is 0.12 and 0.27, respectively, in units of $e^2/\varepsilon \ell$, 
which are roughly consistent with the
result 0.20 obtained from the exact diagonalizations in the
system of spherical geometry\cite{Onod}.

In Fig.~2 (a) we show the ground state pair correlation function
defined in the following equation:
\begin{equation}
g({\bf r}) \equiv \frac{L_x L_y}{N_e(N_e-1)}\langle 
\Psi | \sum_{i\neq j} \delta({\bf r}+{\bf R}_i-{\bf R}_j)|\Psi
\rangle,
\end{equation}
where $|\Psi\rangle$ is the ground state.
The correlation function is circularly symmetric and
short range power low exponent is two as shown in Fig.~3. 
These results are also consistent with gapless liquid ground state.

\begin{figure}
\epsfxsize=80mm \epsffile{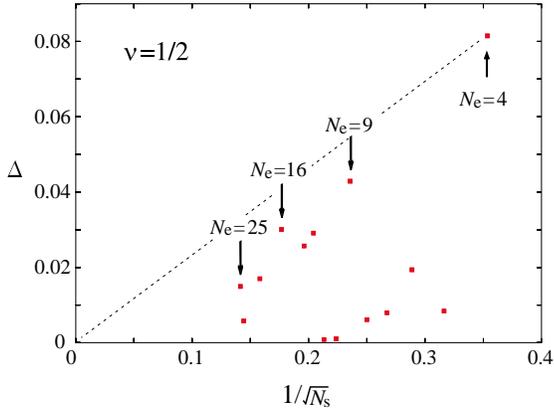}
\caption{
Size dependence of the lowest excitation gap at $\nu=1/2$ for 
various number of electrons $N_e$ in square unit cell. 
$N_s$ is the number of one-particle states in the unit cell.
The dashed line is a guide for eyes.
}
\end{figure}

\subsection{Incompressible liquids at $\nu = n/(1+2n)$}
When we decrease filling factor from 1/2, fractional 
quantum Hall effects are experimentally observed at various 
fractional fillings\cite{FQHE,FQHE2}.
In particular, the primary series of $\nu = n/(1+2n)$ with
$n=1,2,3,4,5$ is well known for its large excitation gap.
Here, we confirm the existence of excitation gap at $\nu = n/(1+2n)$
by systematically calculating the lowest excitation gap for 
filling factors between  $\nu=0.3$ and 0.5.
The obtained result for systems with 12, 14, 15, 16, 18, 20, 
24, and 25 electrons are shown in Fig. 4. 
As is clearly seen, large excitation gap is obtained 
for $\nu = n/(1+2n)$. The size of the gap  
decreases with increasing $n$, and its dependence is 
consistent with experiments. 
Based on the composite fermion theory, the effective mass
is estimated from the excitation gap.
The estimated value of $1/m^*$ in units of $e^2/\varepsilon \ell$ 
is around 0.23, which is almost the same to the value $\sim 0.2$
determined at $\nu=1/2$.
We mention that the composite fermion picture also 
predicts small excitation gap in finite system at $\nu=4/11$\cite{CFLiq},
which is consistent with the present result shown in Fig.~4.

In order to confirm liquid ground state at $\nu = n/(1+2n)$,
we next see the ground state pair correlation functions.
The results for $\nu=5/11, 2/5$, and $1/3$ are shown in 
Figs.~2 (b), (c) and (f), respectively.
As is seen, the correlation functions are circularly symmetric,
which confirms liquid ground state.

\begin{figure}
\epsfxsize=80mm \epsffile{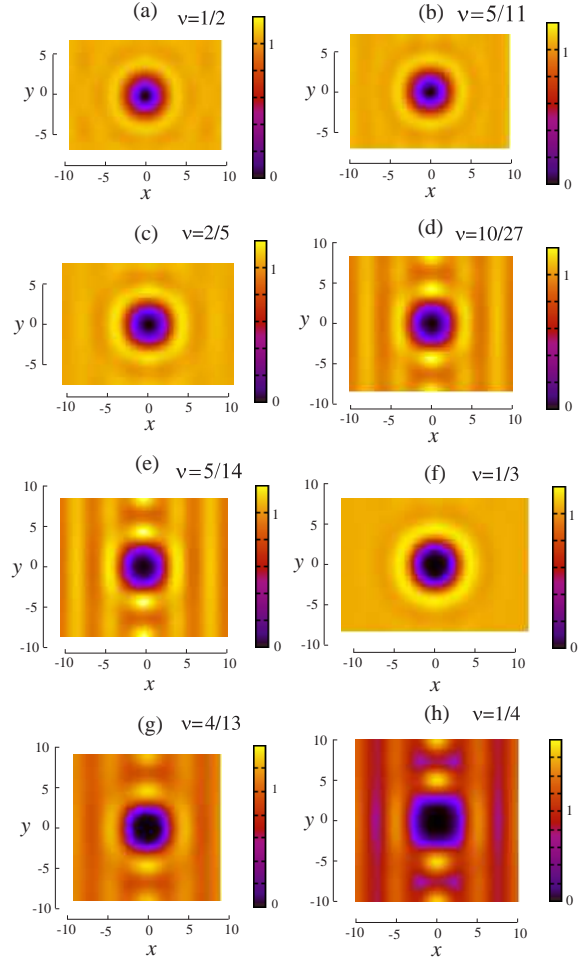}
\caption{
Ground state pair correlation functions in the lowest Landau level 
for various $\nu$. $N_e=20$ for $\nu=$ 
1/2, 5/11, 2/5, 10/27, 5/14, and 1/3. $N_e=16$ for 
$\nu=$ 4/13 and 1/4.}
\end{figure}

\begin{figure}
\epsfxsize=80mm \epsffile{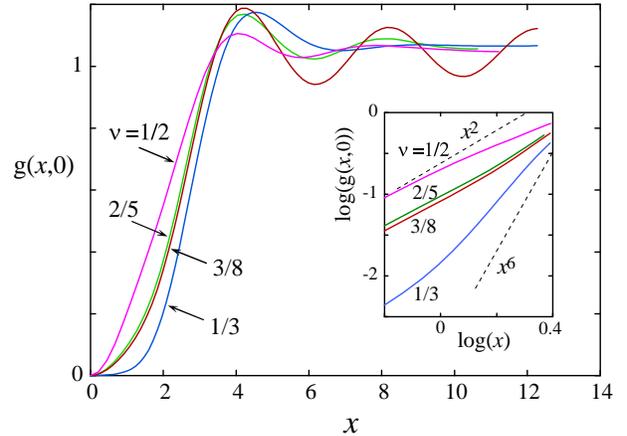}
\caption{
Pair correlation functions in the lowest Landau level for 
various $\nu$ near half filling. Inset shows logarithmic 
plot near the origin.
}
\end{figure}

\begin{figure}
\epsfxsize=80mm \epsffile{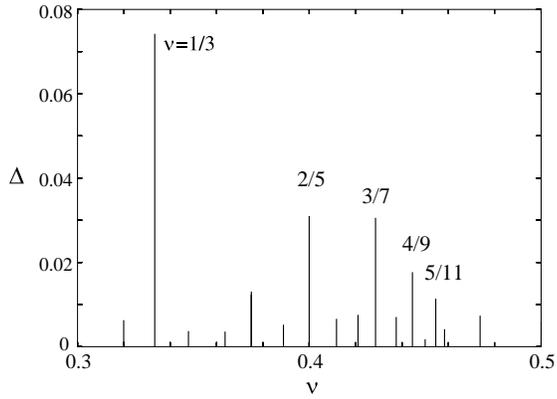}
\caption{
The lowest excitation gap at various $\nu$.
The number of electrons $N_e$ is between 12 and 25. 
}
\end{figure}

\begin{figure}
\epsfxsize=80mm \epsffile{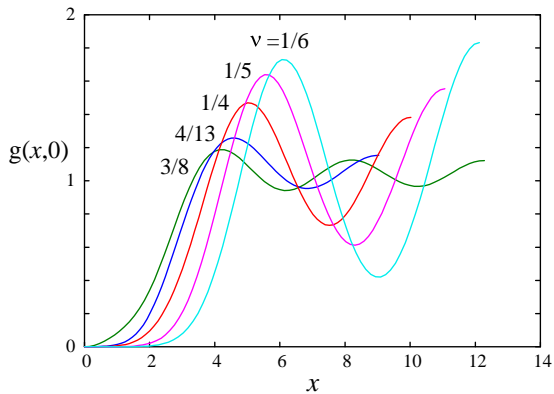}
\caption{
Ground state pair correlation functions in the lowest Landau level 
at various $\nu$. $N_e=24$ for $\nu=$ 3/8. $N_e=16$ for $\nu=$ 4/13 and 
1/4. $N_e=12$ for $\nu=$ 1/5 and 1/6.}
\end{figure}

\subsection{Weak stripes below $\nu \sim 0.42$}
We next consider almost gapless ground state between the
incompressible liquid states.
In Figs.~2 (d), (e), (g), and (h), the pair correlation functions 
at $\nu=10/27$, $5/14$, $4/13$ and $1/4$ are shown. 
We find weak stripe correlations.
Similar stripes are observed
for $\nu  \stackrel{<}{_\sim} 10/24$ except around the 
incompressible states at $\nu=2/5$ and $1/3$. 
The stripe correlations are gradually enhanced 
with decreasing $\nu$ as shown in Fig.~5.
We think the gradual enhancement in the stripe
correlations are due to the instability to Wigner crystal,
and quantum fluctuations partially melt
Wigner crystal to form stripe ground state. 
In fact, short range correlations below $r \sim 5 \ell$ 
are almost the same to that of Wigner crystal. 
This is clearly different from
the stripe state observed in higher Landau levels, where
anisotropic resistivity is experimentally observed.
The difference from the stripe state in higher Landau levels
is clearly shown in \S 3.4.
We note that the present results are different from those
obtained in recent studies around $\nu=3/8$, where
stripe state with much longer period of $10\ell$\cite{CFSt} and
bubble state around $\nu=3/8$\cite{CFBu} are predicted based on 
the composite fermion theory.
The present DMRG results show that the period of the stripes at $\nu=3/8$ 
is about $4\ell$ as shown in Fig.~3 and 
the stripe state is stable in wide range below $\nu\sim 0.42$ 
except around incompressible state at $\nu=2/5$ and $1/3$.
Detailed study at $\nu=3/8$ with small size dependence 
of the stripes will be published elsewhere.

\subsection{Laughlin state and stripe order at $\nu=1/5$}
The stripe correlation in the ground state 
is enhanced with decreasing $\nu$.
The correlation function is anisotropic even at $\nu=1/5$
as shown in Fig.~6 (b), 
and it is not clear whether there exist transition to 
incompressible state around $\nu\sim 1/5$.
In order to identify the ground state at 
$\nu=1/5$, we introduce Haldane's pseudo potentials 
$V_m$ between the electrons\cite{pseudo}. 
It is shown that the Laughlin state at $\nu=1/5$ is an exact 
ground state for strong short range repulsive interaction 
consisting of only $V_1$ and $V_3$, which act only between 
electron pairs whose 
relative angular momentum $m$ is 1 and 3, respectively.
Thus we can see how the ground state connects to Laughlin state
with increasing $V_1$ and $V_3$ from its values
of the pure Coulomb interaction. If there exist no phase transition, 
the ground state is in the same phase characterized by Laughlin state.

In order to clarify whether there is a phase transition, we 
calculate the low energy excitations as a function of $V_3$. 
Since $V_1$ of pure Coulomb interaction
is already large in the lowest Landau level with a small probability of 
finding electron pairs whose relative angular momentum is 1 for $\nu < 1/3$, 
we change only $V_3$. 

Figure~7 shows the $V_3$-dependence of the ground state energy 
and the first excited state energy at $\nu=1/5$ with 12 electrons. 
With increasing $V_3$ from $\delta V_3=0$, the ground state energy 
monotonically changes and the excitation gap continuously increases.
The ground state correlation function at $\delta V_3 = 0.24$ shown in 
Fig.~6 (a) is
circularly symmetric and it is well characterized by Laughlin state. 
Since no phase transition is detected,
the ground state of pure Coulomb 
interaction belongs to the same phase of Laughlin state. 

When we decrease $V_3$ from $\delta V_3=0$, however, 
significant change occurs in the ground state. 
As shown in Fig. 7, the slope of the ground state energy 
clearly changes at $\delta V_3 = -0.04$. 
This change in the ground state energy reflects the
qualitative change in short range correlation functions. 
As is shown in Fig. 8, the shoulder structure at $x\sim 3\ell$ 
develops below $\delta V_3 \sim -0.04$ and it makes clear anisotropy 
around $r \sim 4\ell$.
The shape of the correlation function for $\delta V_3 < -0.04$
is almost the same to that realized in higher 
Landau levels and clearly different from the stripe structure 
of pure Coulomb interaction in the lowest Landau level around 
$\nu=1/5$.

\begin{figure}[t]
\epsfxsize=80mm \epsffile{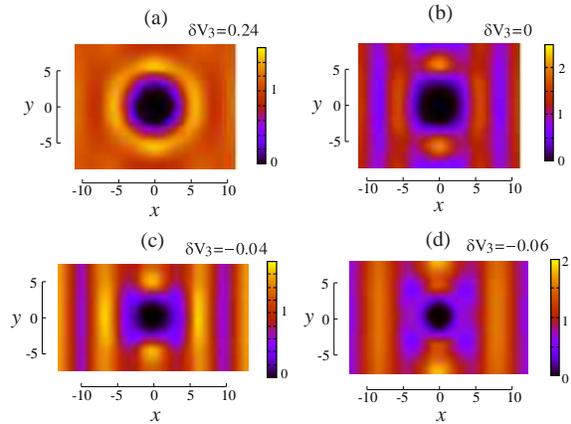}
\caption{
Ground state pair correlation functions 
at $\nu=1/5$ for various $\delta V_3$. $N_e=12$. 
$\delta V_3=0$ corresponds to pure Coulomb interaction.
}
\end{figure}

\begin{figure}
\epsfxsize=80mm \epsffile{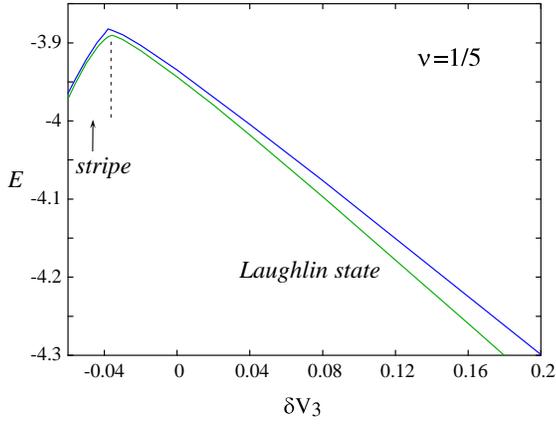}
\caption{$V_3$ dependence of the ground state and first excited 
state energies at $\nu=1/5$ in the lowest Landau level.
$N_e=12$. $\delta V_3=0$ corresponds to pure Coulomb interaction.
}
\end{figure}

\begin{figure}
\epsfxsize=80mm \epsffile{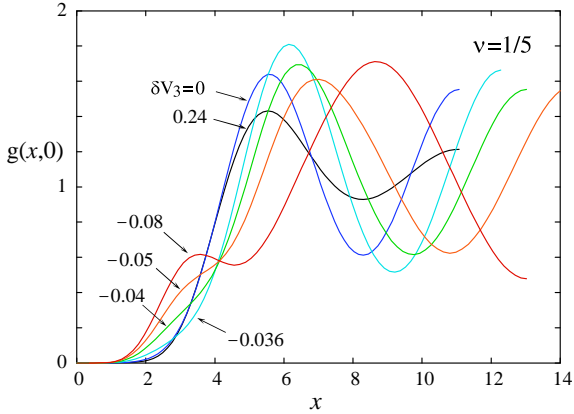}
\caption{
Ground state pair correlation functions on the $x$-axis at $\nu=1/5$ 
for various $\delta V_3$. $N_e=12$.
$\delta V_3=0$ corresponds to pure Coulomb interaction.
}
\end{figure}

\subsection{Transition to Wigner crystal}
In the limit of low filling, the magnetic length $\ell$ 
becomes negligible 
compared with the mean distance between the electrons. Since $\ell$ 
is the length scale
of one particle wave function, the electrons can be
treated as classical point charges, and the ground state is 
expected to be Wigner crystal. 
In order to confirm the existence of Wigner crystal at finite $\nu$,
we next see pair correlation functions in the 
ground state at low fillings. 
The ground state pair correlation functions below $\nu=3/8$ 
in Figs.~5 and 9 (a) show that the stripe structure remains stable down 
to $\nu\sim 1/6$, 
although it seems to approach hexagonal structure of
Wigner crystal. 

At $\nu=1/6$ the Wigner crystal is realized in an excited state close to
the ground state. The correlation function of the Wigner crystal 
and the energy difference from the ground state
are shown in Fig.~9 (b) and (c).
The energy difference almost linearly decreases with decreasing $\nu$ 
except at $\nu=1/5$, where the ground state is characterized by
Laughlin state. Linear extrapolation on $\nu$ shows the crossing of the 
two energy levels at around $\nu=1/7$. 
Since the momentum of the two states are different, we expect first order 
transition to the Wigner crystal at $\nu_c\sim 1/7$.
The value of the critical filling $\nu_c$ is consistent with previous 
studies\cite{Lam,Kun},
although exact diagonalizations of 6-electrons predict second order
or weak first order transition\cite{Kun}.

\begin{figure}[t]
\epsfxsize=80mm \epsffile{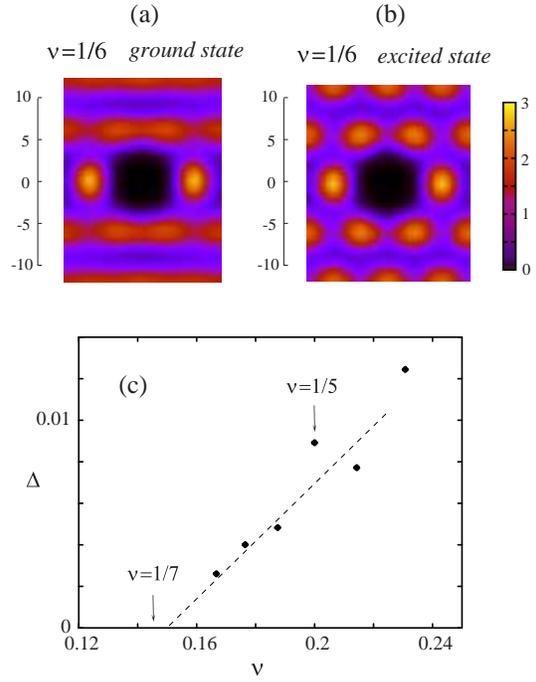}
\caption{
(a) The ground state pair correlation function at $\nu=1/6$.
(b) The pair correlation function of Wigner crystal at $\nu=1/6$.
(c) Energy difference between the ground state and Wigner crystal
at low fillings. $N_e=12$. The dashed line is a guide for eyes. 
}
\end{figure}

\subsection{Phase diagram of $N=0$ Landau level}
The present DMRG calculation shows the existence of four different 
ground states:
the compressible liquid around $\nu=1/2$, the incompressible liquid
at $\nu=n/(2n+1)$ and $1/5$, the weak stripe state below $\nu\sim 0.42$,
and Wigner crystal below $\nu_c \sim 1/7$. From the above results, 
the ground state phase diagram is determined as shown in Fig.~10.
Precise critical fillings and the nature of the transitions are not 
clear for the boundary between the liquid states and the weak stripe 
state, but the transition to Wigner crystal is expected to be 
first order.
The weak stripe structure found below $\nu\sim 0.42$ has clear
oscillations along the stripes and this is clearly different
from the stripe state realized in higher Landau levels. 
The ground state at $\nu=1/5$ continuously connects to the 
Laughlin state realized in the limit of large $V_1$ and $V_3$,
while the size of the excitation gap is vary small compared with 
the gap at $\nu=1/3$. The pair correlation function at $\nu=1/5$ 
is strongly modified from circularly symmetric one, because 
$V_3$ of the Coulomb interaction is located close to the
boundary between Laughlin state and stripe state.

\begin{figure}
\epsfxsize=80mm \epsffile{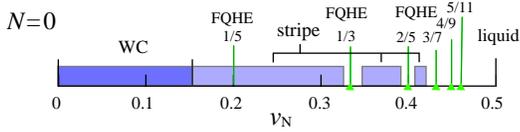}
\caption{
Ground state phase diagram of the lowest Landau level
determined by the DMRG method.
}
\end{figure}

\subsection{Pairing state at half filling in $N=1$ Landau level}
We next investigate the ground state in the second lowest Landau level. 
At $\nu_N=1/2$, fractional quantum Hall effect has been experimentally
observed\cite{exp52}. To explain the fractional quantization, $p$-wave
pairing formation of electrons has been proposed\cite{RezHal,grei,Moor}. 
In particular, recent studies based on exact diagonalizations suggest
Paffiann state of triplet spin pair\cite{morf,Rez52}. 

In the present study we introduce Haldane's pseudo-potential $V_1$
in the second lowest Landau level. We calculate the excitation energies
and the pair correlation functions at various $V_1$.
In Fig.~11 we show the total energy of the ground state and several 
low energy excited states as a function of $\delta V_1$. 
We find two characteristic $\delta V_1$ where structure of the low energy 
spectrum changes. 
One is located slightly below $\delta V_1=0$ and the other is 
around $\delta V_1=0.03$ as shown in the figure by arrows.
Below $\delta V_1=0$ the ground state is almost degenerate and
the pair correlation function is characterized by
clear stripes as shown in Fig.~12 (c).  
For $V_1>0.03$, the ground state and low energy excited states
is almost independent of $V_1$. The correlation function is 
circularly symmetric with weak oscillations as shown 
in Fig.~12 (a), and it is similar 
to that observed at $\nu=1/2$ in the lowest Landau level, where 
excitation gap vanishes in the thermodynamic limit.
Since sufficiently large $\delta V_1$ effectively maps the system onto the 
lowest Landau level, we expect that the ground state is compressible 
liquid for $\delta V_1>0.03$.

Between $\delta V_1=0.0$ and $0.03$,  the ground state is clearly 
different from the stripe state. 
The correlation function clearly decays at long distance as shown in 
Fig.~12 (b) and the degeneracy in the 
ground state is lifted with a finite excitation gap of order 0.01.
Compared with the compressible liquid state at $\delta V_1>0.03$,
the correlation function is 
different at short distance as shown in Fig.~12 (d), which shows 
the difference in the correlation function 
between $\delta V_1=0.01$ and $0.06$: 
$g(r)_{\delta V_1=0.01}-g(r)_{\delta V_1=0.06}$.
In this figure, we can see that electrons are attracted around the origin
$r\sim 1.5$ consistent with the pair formation. 

\begin{figure}
\epsfxsize=80mm \epsffile{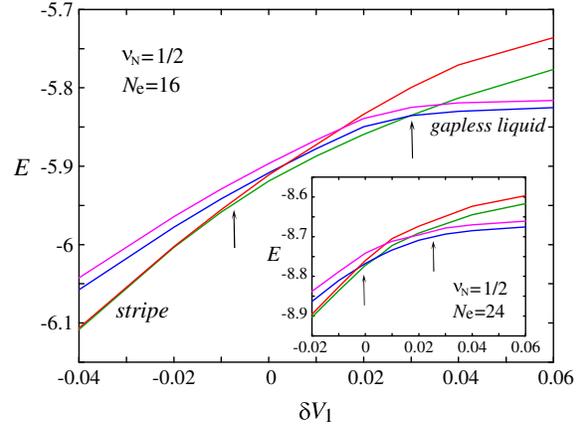}
\caption{Low energy spectrum at half filling in the
second lowest Landau level. $N_e=16$. Inset shows the results 
for $N_e=24$. $\delta V_1=0$ corresponds to pure Coulomb interaction.
}
\end{figure}

\begin{figure}
\epsfxsize=80mm \epsffile{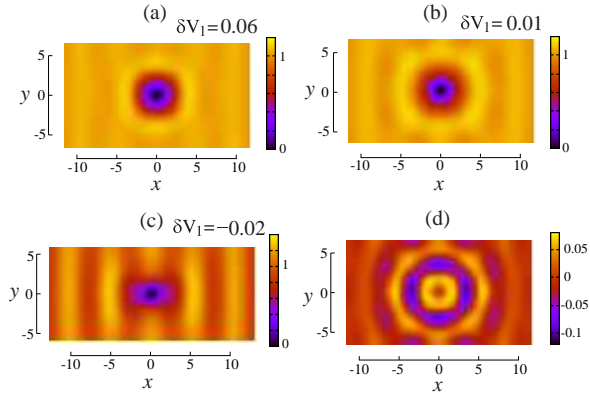}
\caption{The ground state pair correlation functions
in guiding center coordinates at half filling in the 
second lowest Landau level at $\delta V_1=0.06$ (a), 
$0.01$ (b), and $-0.02$ (c). 
Figure (d) shows the difference in the correlation function 
between $\delta V_1=0.01$ and $\delta V_1=0.06$, 
$g(r)_{\delta V_1=0.01}-g(r)_{\delta V_1=0.06}$. $N_e=24$. 
}
\end{figure}

\subsection{Stripes away from half filling in $N=1$ Landau level}
We next investigate the ground state away from half filling.
The pair correlation functions for $\nu=$ 4/9, 2/5, 4/11, and 1/3 with 
16 electrons in the unit cell are presented in Fig.~13. 
These results clearly show the existence of the stripe correlations 
even in the second lowest Landau level near half filling. 
In order to see detailed structure of the stripes
we show the correlation functions on the $x$ and $y$ axes in Fig.~14. 
As is shown, the correlation function 
on the $y$-axis at $\nu=$ 4/9 and 2/5
shows monotonic increase from the origin and
it becomes almost constant at a long distance.
This is the same structure of the stripes observed in higher
Landau levels near half filling as shown in Fig.~15.
Compared with the stripes in $N=2$ Landau level,
the amplitude of the oscillations on the $x$-axis is about 
50\% smaller and their period is 30\% shorter.
This means the strips in $N=1$ Landau level is likely to
be disturbed by residual potential fluctuations 
and temperatures. We think 
this may be a reason why anisotropic resistivity
is not clearly observed in the second lowest Landau level.

With further decreasing $\nu$, character of the stripes changes at 
$\nu\sim 4/11$. This is clearly shown in the correlation functions 
along the stripes in Fig.~14. 
The correlation function for $\nu<4/11$ shows weak oscillations
along the stripes, which is similar to the stripe structure in 
the lowest Landau level. 
The oscillations on the stripes are enhanced with decreasing $\nu$
and the ground state becomes almost the same to that in the 
lowest Landau level except at $\nu=1/3$.
At low fillings $\nu<1/4$, the correlation functions 
in $N$=0 and 1 Landau levels are almost identical with each other.
The similarity of the ground state between the two Landau levels at
$\nu=1/4$ has been shown by previous calculations\cite{MorAm,Ono2}.

\begin{figure}[t]
\epsfxsize=80mm \epsffile{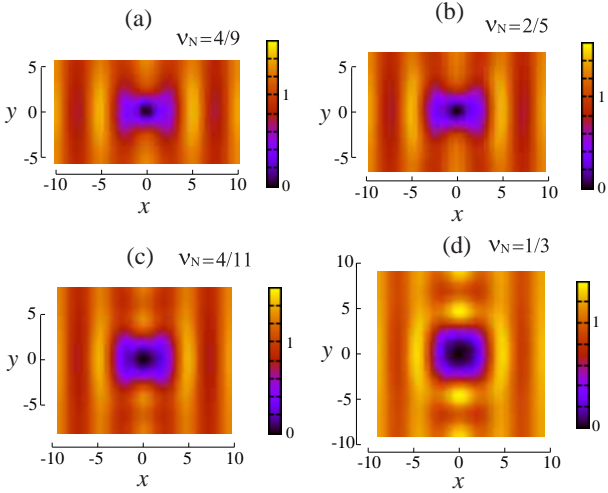}
\caption{The ground state pair correlation functions
in guiding center coordinates
in the second lowest Landau level near half filling.
$N_e=16$. 
}
\end{figure}

\begin{figure}[t]
\epsfxsize=80mm \epsffile{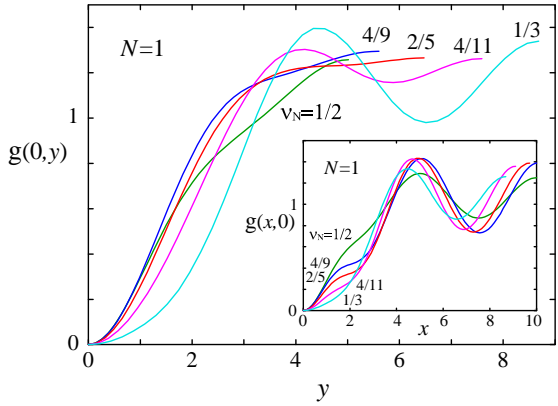}
\caption{
The ground state pair correlation functions on the $y$-axis
in guiding center coordinates
in the second lowest Landau level. 
Inset shows the results on the $x$-axis. $N_e=16$.
}
\end{figure}

\begin{figure}
\epsfxsize=80mm \epsffile{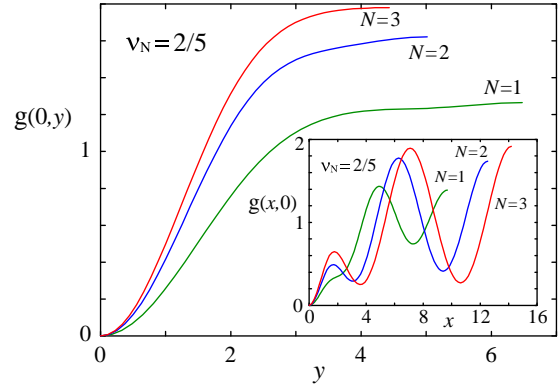}
\caption{
The ground state pair correlation functions on the $y$-axis
in guiding center coordinates
in $N$ = 1, 2, and 3 Landau levels at $\nu_N=2/5$. 
Inset shows the results on the $x$-axis.  $N_e=16$.
}
\end{figure}

\begin{figure}
\epsfxsize=80mm \epsffile{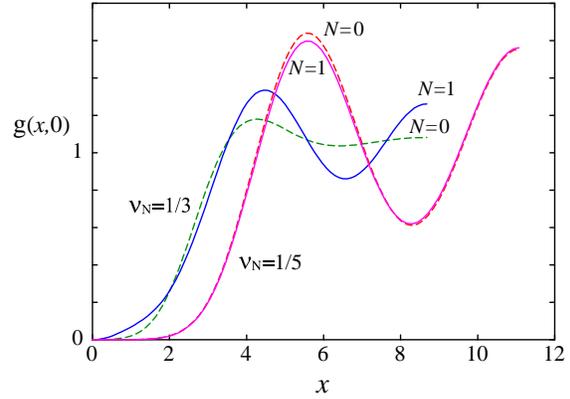}
\caption{
The ground state pair correlation functions at 
$\nu_N=1/3$ and $1/5$ in the lowest and the second lowest 
Landau levels.
}
\end{figure}

At $\nu=1/3$, the ground state correlation function is
clearly different from that of the lowest Landau level.
Compared with Laughlin state in the lowest Landau level,
short range correlation below $r\sim 2\ell$ is enhanced 
and clear oscillations exist at long distance as shown in Fig.~16.
This is due to the large reduction of short range interaction
$V_1$, which stabilizes Laughlin state.

In order to see the effect of $V_1$ on the ground state,
we show in Fig.~17 the energies of the ground state and excited state 
as a function of $V_1$. 
The excitation gap monotonically increases with increasing
$V_1$ from $\delta V_1=0$ and no phase transition is
detected. Since the ground state at large $V_1$ is Laughlin
state as shown in Fig.~18 (c), where the correlation functions 
at $\delta V_1=0.08$ is almost identical to that in the lowest 
Landau level, the present result shows the ground state at 
$\delta V_1=0$ continuously connects to Laughlin state.
However, the excitation gap is very small at $\delta V_1=0$ 
compared with the gap in the lowest Landau level and slight 
decrease in $V_1$ leads to the transition to stripe state as shown 
in Fig.~17. The correlation function of the stripe state at 
$\delta V_1=-0.02$ is shown in Fig.~18 (b), which is similar to the
stripe state near half filling. With further decreasing $V_1$,
first order transition to two-electron bubble phase occurs 
at $\delta V_1\sim -0.04$.
This is consistent with the results in higher Landau levels,
where bubble state is realized at $\nu_N=1/3$ in $N=2$
Landau level\cite{Shib,Kou3,Rez2,DY}.

In contrast to the ground state at $\nu_N=1/3$, 
the correlation function at $\nu_N=1/5$ is almost 
the same to that in the lowest Landau level as shown in Fig.~16. 
This is not surprising because 
$V_3$ of the second lowest Landau level is larger than that 
of the lowest Landau level.
Since Laughlin state at $\nu_N=1/5$ is stabilized by $V_1$ and $V_3$, 
large $V_3$ compared with other $V_m$ 
stabilizes Laughlin state similarly to the lowest Landau level,
although circularly symmetric correlation function is 
strongly modified by the presence of 
other pseudo-potentials $V_m$ of Coulomb interaction.

\begin{figure}[t]
\epsfxsize=80mm \epsffile{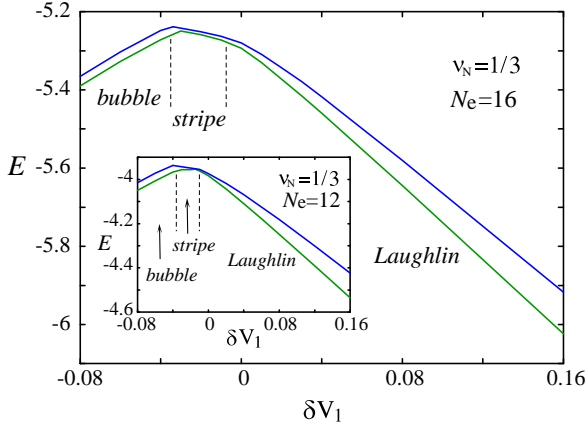}
\caption{
The ground state and first excited state energies
in the second lowest Landau level at $\nu_N=1/3$
with $N_e=16$. 
Inset shows the results for $N_e=12$.
$\delta V_1=0$ corresponds to pure Coulomb interaction.
}
\end{figure}

\begin{figure}[t]
\epsfxsize=85mm \epsffile{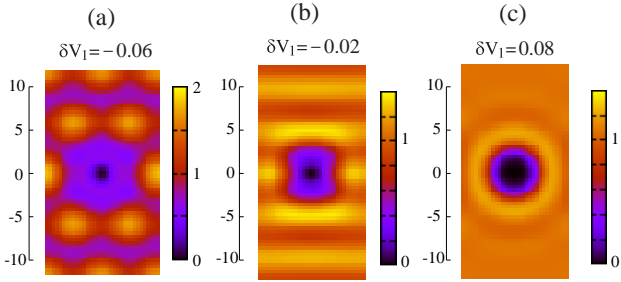}
\caption{
The ground state pair correlation functions 
in guiding center coordinates
in the second lowest Landau level at $\nu_N=1/3$
with $N_e=16$. 
}
\end{figure}

\subsection{Transition to Wigner crystal}
At sufficiently small $\nu_N$, the Wigner crystal is realized 
in the ground state even in high Landau levels, because 
electrons can be treated as point particles when 
mean distance between the electrons becomes much longer than
typical length scale of the one-particle wave function.
In order to study the transition to Wigner crystal in the
second lowest Landau level, we calculate the energy of 
Wigner crystal and compare with the ground state energy.
The energy difference between the two states and 
the correlation functions of the guiding center are shown in 
Figs.~19 and 20, respectively.
The energy difference between the two states monotonically decreases 
except at $\nu_N=1/5$, where the ground state continuously connects to 
Laughlin state. The singular behavior at $\nu_N=1/5$ is due to the decrease
in the ground state energy.
The linear extrapolation of the energy difference 
with respect to $\nu_N$ shows first order transition to Wigner crystal
at $\nu_N\sim 1/7$.

\begin{figure}[t]
\epsfxsize=80mm \epsffile{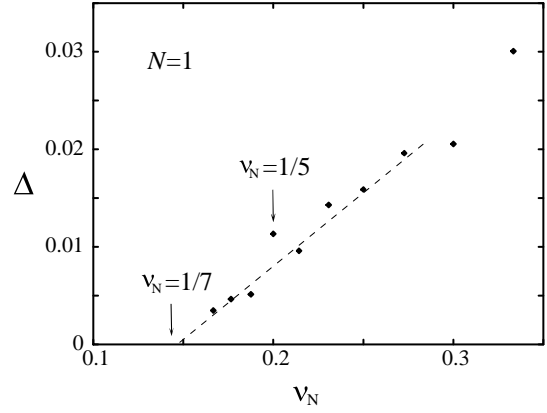}
\caption{
Energy difference between Wigner crystal and the ground state.
$N_e=12$. The dashed line is a guide for eyes.
}
\end{figure}

\begin{figure}[t]
\epsfxsize=80mm \epsffile{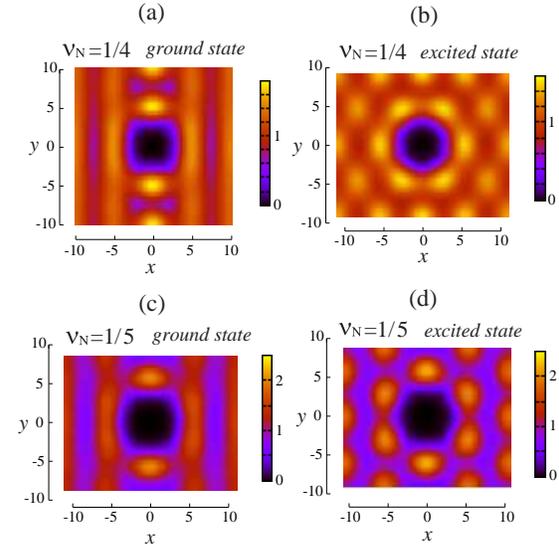}
\caption{
Pair correlation functions in guiding center coordinates
in the second lowest Landau level. 
(a) The ground state at $\nu_N=1/4$ with 16 electrons. 
(b) Wigner crystal at $\nu_N=1/4$ with 16 electrons obtained 
at a different total momentum from the ground state. 
(c) The ground state at $\nu_N=1/5$ with 12 electrons. 
(d) Wigner crystal at $\nu_N=1/5$ with 12 electrons.
}
\end{figure}

\subsection{Phase diagram of $N=1$ Landau level}
The ground state phase diagram of the second lowest Landau level
is shown in Fig.~21. 
At half filling, the ground state pair correlation function suggests
pairing state. This ground state is different from both the stripe 
state in higher Landau levels and the compressible liquid state in 
the lowest Landau level.
With decreasing $\nu_N$, however, stripe state similar to that in 
higher Landau levels is realized with small amplitude of the 
stripes (stripe I),
which is 50\% smaller than that in $N=2$ Landau level.
With further decreasing $\nu_N$, clear oscillations 
along the stripes appear below $\nu \sim 4/11$ (stripe II). 
This structure is similar to the stripes 
in the lowest Landau level.
At $\nu_N=1/3$, the ground state belongs to the same phase of 
Laughlin state in the lowest Landau level, but correlation function 
shows clear oscillations at long distance.
This is due to the proximity to the boundary between Laughlin state 
and stripe state.
Below $\nu_N\sim 1/4$, the correlation functions are almost identical
to that in the lowest Landau level, and thus the ground state is almost 
the same between the two Landau levels.
The ground state at $\nu_N=1/5$ is expected to be in the same phase
of Laughlin state. The first order transition to Wigner crystal is 
expected to occur at $\nu_N\sim 1/7$.

\begin{figure}[t]
\epsfxsize=80mm \epsffile{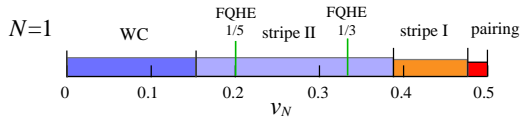}
\caption{
Ground state phase diagram of the second lowest Landau level
determined by the DMRG method.
}
\end{figure}

\section{Summary and Discussion}

In the present paper we have studied ground state pair correlation 
functions and
low energy excitations of 2D electrons in the lowest and the second lowest
Landau levels by using DMRG method. 
We have assumed completely spin polarized state and neglected
Landau level mixing. The obtained results  in the lowest Landau level
confirmed the existence of compressible liquid at $\nu=1/2$, incompressible 
liquid at $\nu=n/(1+2n)$ and $1/5$, and Wigner crystal below $\nu\sim
1/7$. We have found new stripe
state between $\nu\sim 0.42$ and $1/7$, whose correlation function shows 
oscillations along the stripes, which is different from the stripes 
in higher Landau levels. In the second lowest Landau level, pairing
state at $\nu_N=1/2$ has been confirmed. 
The ground state is close to the phase boundary between the pairing
state and stripe state, and decrease in $V_1$
of about 2\% is enough for the transition to the stripe state.
Between $\nu_N\sim 0.47$ and $0.38$ we have found stripe state similar to 
that in higher Landau levels with small amplitude of stripes.
We have also found stripe state similar to that in the lowest 
Landau level between $\nu_N\sim 0.38$ and $1/7$.
At low fillings $\nu_N < 1/3$, the ground state is
almost the same to that in the lowest Landau level.
This is due to the fact that $V_3$ of the second lowest 
Landau level, which is the dominant interaction for $\nu_N < 1/3$,
is slightly increased from that in the lowest Landau level,
and this increase stabilizes 
the ground state realized in the lowest Landau level.

We have also shown the existence of incompressible liquids
at $\nu_N=1/3$ and $1/5$ in the second lowest Landau level.
The size of the gap at $\nu_N=1/3$ is very small
compared with that in the lowest Landau level and
slight decrease in $V_1$ of about 0.01 leads to the transition to 
stripe state. In actual systems, the wave function has finite 
width in the direction perpendicular to the two-dimensional plane.
Since finite width reduces $V_1$ from the value of
ideal two dimensional system, incompressible liquid at $\nu_N=1/3$
will not be realized in systems of wide quantum well.
Our estimate shows the critical width is about $4\ell$ for
rigid density distribution in the direction perpendicular to 
the two-dimensional plane in square potential well. 

Recent experiment on the second lowest Landau level
shows that there exist reentrant integer quantum Hall 
states around $\nu_N=0.42$ and $\nu_N=0.25$\cite{exp1}.
Our results suggest reentrant phase around $\nu_N=0.42$
corresponds to the stripe phase near half filling, although 
the mechanism of the integer quantum Hall effect is still not clear.
Since the amplitude of the stripes is small compared with 
that of higher Landau levels, it is expected that
weak random potentials disturb stripe correlations and 
insulating state pined by residual potential fluctuations 
is realized before forming stripe state.

Concerning to the reentrant phase around $\nu_N=0.25$, similar 
behavior is expected in the lowest Landau level,
since the ground state obtained in the present calculation
is almost the same between the two Landau levels.
This means the origin of insulating behavior above $\nu=1/5$
in the lowest Landau level is the same to that of the 
reentrant phase around  $\nu_N=0.25$ in the second lowest Landau level.
Since the energy of Wigner crystal is close to the ground state energy, 
it may be possible that random potentials stabilize pined Wigner crystal, 
which is the same mechanism of integer quantum Hall effect
at small $\nu_N<1/7$.

\acknowledgement

The present work is supported by
Grant-in-Aid No.~14540294 and No.~11740184 from JSPS.

\end{document}